\documentclass[runningheads]{llncs}
\usepackage{cite}
\usepackage{amsmath} 
\usepackage{comment}
\usepackage{booktabs, tabularx} 
\usepackage{amsmath, amssymb}
\usepackage{bm,url}
\usepackage{amsmath}
\usepackage{svg}
\usepackage{placeins}
\usepackage{algorithm}
\usepackage{algorithmic}
\usepackage{mathtools}
\usepackage{siunitx}
\usepackage{hyperref}
\usepackage[T1]{fontenc}
\usepackage{graphicx,verbatim}
\usepackage{hyperref}
\usepackage{color}

\begin{document}
\title{VidFuncta: Towards Generalizable Neural Representations for Ultrasound Videos}
\titlerunning{VidFuncta for Ultrasound Videos}

\author{Julia Wolleb\inst{1,2} \and
Florentin Bieder\inst{3} \and
Paul Friedrich\inst{3} \and
Hemant D. Tagare$^*$ \inst{4,5} \and
Xenophon Papademetris$^*$ \inst{1,2,4,5}}

\authorrunning{J. Wolleb et al.}

\institute{Dept. of Biomedical Informatics \& Data Science, Yale University, New Haven, USA \and
Yale Biomedical Imaging Institute, Yale University, New Haven, USA \and
Dept. of Biomedical Engineering, University of Basel, Allschwil, Switzerland \and
Dept. of Radiology \& Biomedical Imaging, Yale University, New Haven, USA \and
Dept. of Biomedical Engineering, Yale University, New Haven, USA 
\email{julia.wolleb@yale.edu}\\
}

\maketitle              
\begin{abstract}
Ultrasound is widely used in clinical care, yet standard deep learning methods often struggle with full video analysis due to non-standardized acquisition and operator bias. We offer a new perspective on ultrasound video analysis through implicit neural representations (INRs). We build on \textit{Functa}, an INR framework in which each image is represented by a modulation vector that conditions a shared neural network. However, its extension to the temporal domain of medical videos remains unexplored. To address this gap, we propose \textit{VidFuncta}, a novel framework that leverages \textit{Functa} to encode variable-length ultrasound videos into compact, time-resolved representations.
 \textit{VidFuncta} disentangles each video into a static video-specific vector and a sequence of time-dependent modulation vectors, capturing both temporal dynamics and dataset-level redundancies.
Our method outperforms 2D and 3D baselines on video reconstruction and enables downstream tasks to directly operate on the learned 1D modulation vectors. We validate \textit{VidFuncta} on three public ultrasound video datasets -- cardiac, lung, and breast -- and evaluate its downstream performance on ejection fraction prediction, B-line detection, and breast lesion classification.
These results highlight the potential of \textit{VidFuncta} as a generalizable and efficient representation framework for ultrasound videos. Our code is publicly available under \url{https://github.com/JuliaWolleb/VidFuncta_public}.


\end{abstract}
%
%

\section{Introduction}

\begin{figure}[t]
    \centering
    \includegraphics[width=\linewidth]{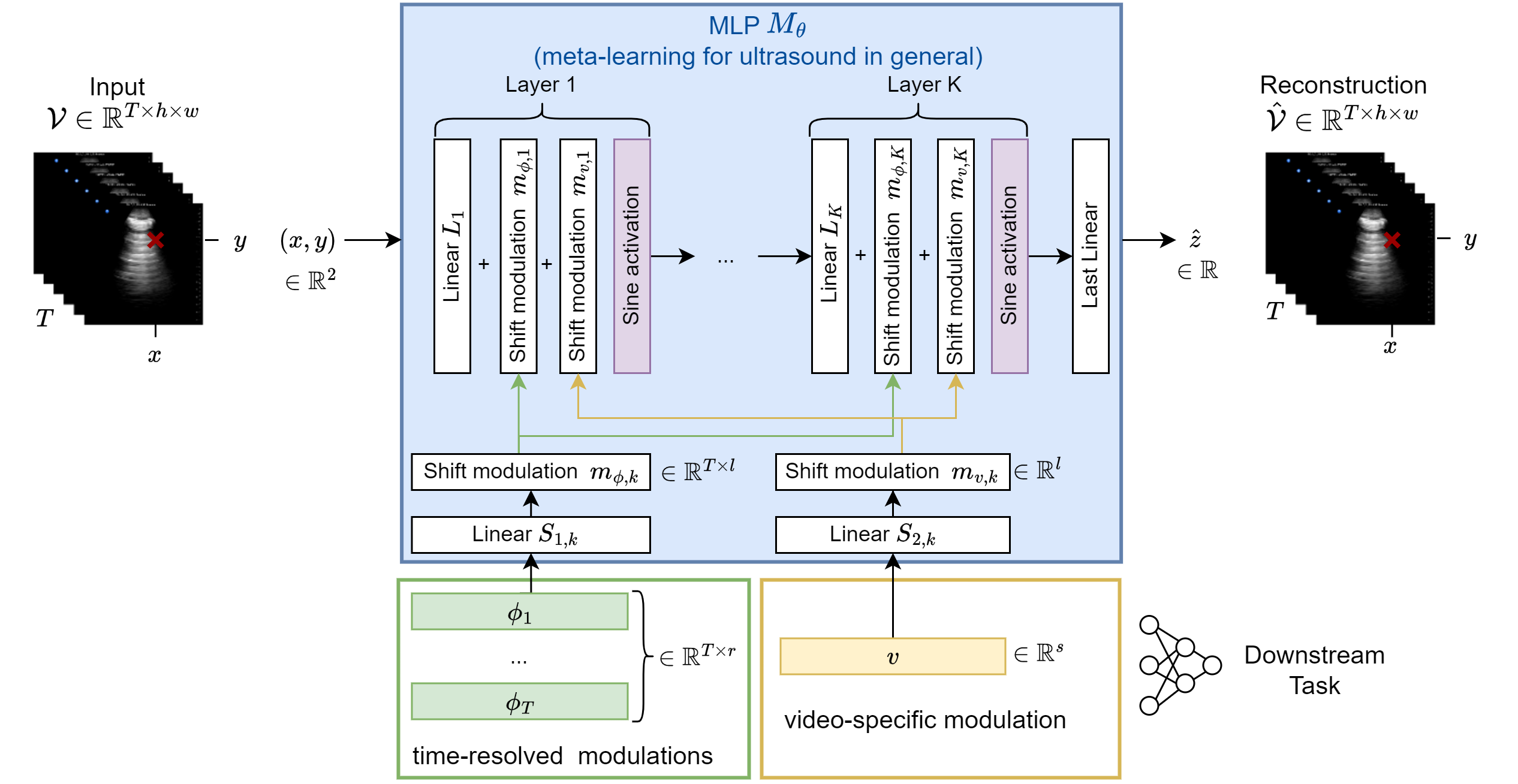}
    \caption{Overview and architecture of our proposed \textit{VidFuncta} framework. For a coordinate $(x,y)$ of a frame, shown as a red cross, the model reconstructs the grayscale value $\hat{z}$. The meta-model in blue captures features shared across the dataset. We compress  an input video into a video-specific modulation vector $v$ (in yellow) capturing features consistent over time, and time-resolved modulation vectors $\{\phi_t\}_{t=1}^T$ (in green) capturing temporal changes. Downstream tasks can be directly applied to these modulations.}
    \label{fig:overview}
\end{figure}

Ultrasound is a fast, affordable, and portable imaging modality, making it especially valuable in emergency care and low-resource settings \cite{stewart2020trends}. Its diagnostic use spans cardiac assessment, lung disease scoring, and tumor evaluation \cite{rumack2023diagnostic}. However, interpretation remains challenging due to non-standardized acquisition, variable image quality, and operator-dependent biases \cite{kim2021artificial}. While deep learning methods have been developed to assist interpretation \cite{wang2021deep}, they often struggle with full-length video analysis due to high redundancy and inconsistencies in acquisition settings \cite{wiedemann2025covid}. To explore an alternative pathway,  we propose a novel approach based on implicit neural representations (INRs).
We build on \textit{Functa} \cite{dupont2022data}, which represents each image as a modulation vector that conditions a shared INR network. This shared network learns a data representation that generalizes across the entire dataset, while the modulation vectors capture image-specific details, enabling efficient compression.
However, this approach is designed for still images and has not been extended to handle video data.
To address this gap, we propose \textit{VidFuncta}, a framework that compresses variable-length ultrasound videos into a single video-specific modulation vector $v$ and a sequence of time-resolved modulation vectors $\{\phi_t\}_{t=1}^T$.  This design leverages redundancy over time and across samples to learn compact, generalizable video representations. An overview is given in Figure~\ref{fig:overview}.
We evaluate our method’s reconstruction performance on three public ultrasound datasets: cardiac \cite{ouyang2019echonet}, lung \cite{asgari2024can}, and breast \cite{lin2022new}.
We explore clinical downstream tasks -- ejection fraction prediction, B-line detection, and breast lesion classification -- on the modulation vectors, which reduces training time and memory usage.\\
\\
\\
\textbf{Related work} 
Deep learning for ultrasound videos has leveraged both 3D \cite{huang2023breast,baloescu2020automated} and 2D convolution-based models \cite{howard2020improving,ouyang2019echonet,wang2021deep}, performing well on  tasks like ejection fraction prediction, lung assessment, and lesion tracking \cite{born2020pocovid,ouyang2019echonet, lin2022new,lucassen2023deep}. Self-supervised methods have also been proposed \cite{Yurong2023Self-supervised, guo2025efficient}.  
However, performance often drops on full videos, leading many to adopt frame-selection strategies for 2D models \cite{born2020pocovid}. These strategies can introduce bias and risk missing critical diagnostic features. Additionally, image quality and domain shifts further impact performance \cite{wiedemann2025covid}.
In this work, we move away from conventional video-based models and explore INRs for ultrasound video analysis, building on \textit{Functa} \cite{dupont2022data, dupont2022coin++}. While INRs have shown promise in video super-resolution \cite{chen2022videoinr}, their use in ultrasound has so far been mostly limited to 3D reconstruction \cite{eid2025rapidvol,gu2022representing}.
 \textit{MedFuncta} \cite{friedrich2025medfuncta} introduced
efficient training for neural fields in medical imaging. However, its input resolution
is limited to  $32 \times 32 \times 32$, restricting direct application to videos treated as
3D volumes. \textit{Spatial Functa} \cite{bauer2023spatial} introduced a patch-based latent structure, allowing for improved downstream performance.
\\
\\
\textbf{Contribution}  
To the best of our knowledge, we are the first to explore \textit{Functa} \cite{dupont2022data} for videos. We propose \textit{VidFuncta} to extract a time-resolved representation of variable length, outperforming both 2D and 3D baselines on image reconstruction and enabling downstream tasks on sequences of 1D modulation vectors. We show that a single model can generalize across multiple ultrasound datasets, exhibits good out-of-distribution performance, and significantly reduces memory and training time of the downstream task compared to convolutional models.

\section{Methods}\label{sec:methods}
\begin{figure}[t]
    \centering
    \includegraphics[width=\linewidth]{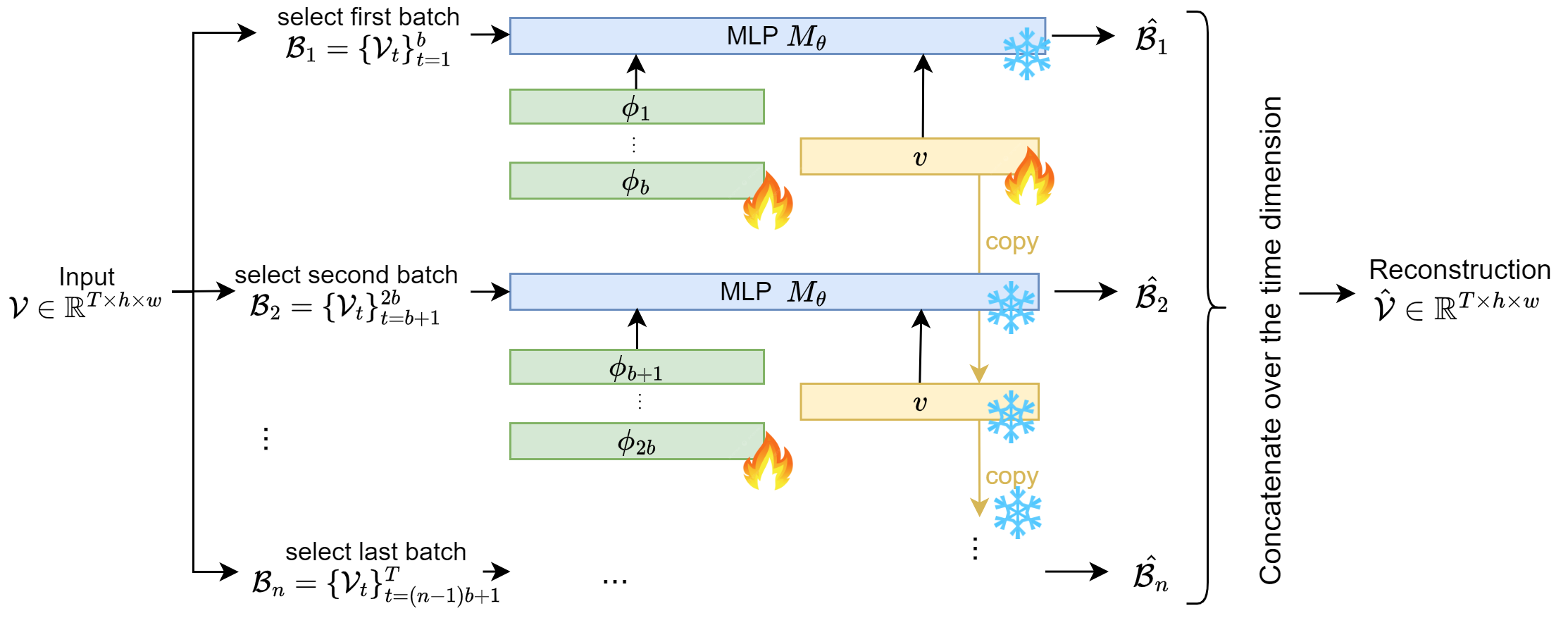}
    \caption{In our autoregressive inference scheme, the video-specific vector $v$ is optimized only in the first batch and frozen afterwards, forcing the modulations $\phi$ to capture temporal changes.}
    \label{fig:inference}
\end{figure}

INRs aim to reconstruct an input signal—in our case, a video $\mathcal{V} \in \mathbb{R}^{T \times h \times w}$—by predicting the grayscale value $z$ at each spatial coordinate $(x, y)$ across frames, where $T$ is the number of frames and $h \times w$ is the size of each frame. We build on \textit{MedFuncta} \cite{friedrich2025medfuncta}, extending its image-level approach to videos by incorporating a time-resolved component into the network architecture, as shown in Figure~\ref{fig:overview}.  This extension is motivated by the need to capture both video features that are stable across time and the dynamic changes between frames, which are critical for accurate ultrasound video modeling.\\
\\
\textbf{Model Architecture}:
The neural network is a multilayer perceptron (MLP) with sinusoidal activation functions \cite{sitzmann2020implicit}.  We adopt a hierarchical design that leverages data redundancy by learning generalizable representations across the ultrasound dataset while conditioning on video- and frame-specific modulation vectors.
At the highest level, the meta-model $M_\theta$ (blue in Figure 1) consists of $K$ linear layers $\{L_k\}_{k=1}^K$, each of dimension $l$, followed by sinusoidal activations. This model, with learnable parameters $\theta$, captures information shared across the entire dataset, such as general anatomical structures.
At the second level, to condition $M_\theta$ on a specific ultrasound video $\mathcal{V}$, we introduce a video-specific modulation vector $v \in \mathbb{R}^{s}$ (yellow in Figure 1). Passing $v$ through a linear layer $S_{2,k}$ produces a shift modulation $m_{v,k}  \in \mathbb{R}^{1 \times l} $, which is added to the output of each layer $L_k$, for $k=\{1,...,K\}$ \cite{perez2018film,dupont2022coin++}. The vector $v$ encodes time-invariant properties such as anatomy, ultrasound gain, and depth.
At the finest level, to capture temporal dynamics within the video, each frame $\{\mathcal{V}_j\}_{j=1}^T$ is associated with a frame-specific modulation vector $\phi_j  \in \mathbb{R}^ r$, forming a sequence $\phi := \{\phi_1, ..., \phi_{T} \}$  (green in Figure 1). A linear projection $S_{1,k}$  maps this sequence to time-resolved shift modulations $m_{\phi,k} \in \mathbb{R}^{T \times l}$, which are also added to the output of $L_k \, \forall k$. \\
\\
\textbf{Model Training}: 
Due to memory constraints, we load one video $\mathcal{V}$ at a time during training, and randomly sample $b$ frames to form a batch $\mathcal{B} \in \mathbb{R}^{b \times h \times w}$.
 The reconstruction loss for the frame at timepoint $t$ is defined as
\begin{equation}
    \mathcal{L}_{MSE,t} =\frac{1}{N} \sum_{i=1}^N \lVert M_{\theta, v, \phi_t}(x_{i},y_{i} ) - z_i \rVert _2^2,
\end{equation}
where $N$ is the number of sampled coordinates per frame and $z_i$ the true grayscale value  at $(x_i,y_i)$.
Following Friedrich et al. \cite{friedrich2025medfuncta}, we adopt a meta-learning strategy with an outer loop to optimize parameters $\theta$ of $M_\theta$, and an inner loop of $G$ steps to optimize the modulation vectors $v$ and $\{\phi_t\}_{t=1}^b$. This process is described in Algorithm 1.
\\
\begin{algorithm}[t] \label{algo_train}
\caption{Optimization Procedure for  \textit{VidFuncta} }
\begin{algorithmic}[1]
\STATE \textbf{Input:} Video $\mathcal{V}$, number of frames $b$, inner loop steps $G$, learning rates $\gamma_1$ and $\gamma_2$ \\
\STATE \textbf{Output:} Trained meta-model $M_\theta$\\
\FORALL{training iterations}
    \STATE $\mathcal{V} \gets$ video loaded from training set \\
     \STATE $\mathcal{B} \gets \text{sample } b \text{ frames from } \mathcal{V}$\\
      \STATE  $ v \gets 0$, $\{\phi_t\}_{t=1}^b \gets 0 $ 
      \COMMENT {Zero-initialization of the latent vectors}
      \FOR{$g = 1$ to $G$} 
            \STATE $\phi_t  \gets\phi_t - \gamma_1 \nabla_{\phi_t} \mathcal{L}_{MSE, t}  \quad \forall t \in \{1,...,b\}$
            \STATE $ v \gets v - \gamma_1 \nabla_v \left(\frac{1}{b}\sum_{t=1}^b\mathcal{L}_{MSE,t} \right)$ 
        \ENDFOR
        \STATE $\theta \gets  \theta- \gamma_2 \nabla_\theta \left(\frac{1}{b}\sum_{t=1}^b\mathcal{L}_{MSE,t} \right)$   
\ENDFOR
\RETURN $M_\theta$
\end{algorithmic}
\end{algorithm}
\\
\textbf{Reconstruction During Inference}:
To handle long videos despite memory limitations, we implement an autoregressive reconstruction approach, as shown in Figure~\ref{fig:inference}. We sample the first batch $\mathcal{B}_1$ consisting of the first $b$ frames of each video $\mathcal{V}$.
We freeze the model parameters $\theta$, and run $G$ inner loop steps to optimize $\{
\phi_t\}_{t=1}^b$ and $v$ according to lines 7 to 10 in Algorithm 1. We assume that this initialization is enough to capture video-specific features, such as the shown anatomy, in the vector $v$. We therefore freeze $v$ for all subsequent batches, and only optimize $\{\phi_i\}_{i=(B-1)b+1}^{Bb}$   for all subsequent batches $\mathcal{B}_B$, with $B=2,...,\left\lceil\frac{T}{b}\right\rceil$. 
To reconstruct a batch $\hat{\mathcal{B}}$, we compute
$\{\hat{z_j}=M_{\theta,v,\phi_t}(x_{j},y_{j})\}_{j=1}^{h*w}$ for all desired spatial coordinates for all frames of $\mathcal{B}$. The final reconstruction $\hat{\mathcal{V}}$ is obtained by concatenating the reconstructed batches along the temporal dimension.

\section{Experiments}
We evaluate our approach on three public datasets. The BEDLUS dataset \cite{asgari2024can,lucassen2023deep} contains 2,026 lung ultrasound videos annotated for the presence or absence of B-lines. The EchoNet-Dynamic dataset \cite{ouyang2019echonet} includes 10,030 cardiac videos labeled with ejection fraction values. The Breast Ultrasound Video dataset  \cite{lin2022new} comprises 188 videos, each annotated with a lesion classification as either benign or malignant.
In addition to training on each dataset individually, we also create a \textit{mixed dataset} composed of 188 breast, 190 lung, and 190 cardiac ultrasound videos. All videos are downsampled to a spatial resolution of $112\times 112$ and normalized to values between 0 and 1. We split a $10\%$ test set from each dataset, and perform 5-fold cross-validation on the remaining data.
We use PyTorch version 2.4.1 for model training. The model architecture uses  $K=10$  layers with a hidden dimension of $l=256$.  The video-specific modulation vector $v$ has dimension $s=2048$, and the modulation vectors $\phi$ have dimension $r=512$, resulting in a compression rate of roughly 24. We perform $G=10$ inner-loop adaptation steps with a learning rate of $\gamma_1=0.1$, and set the meta-learning rate to $\gamma_2=\num{0.5e-6}$. All models are trained for 100,000 iterations on a 24GB NVIDIA RTX A5000 GPU, which takes 20 hours per model. All remaining hyperparameters follow the configuration suggested in \cite{friedrich2025medfuncta}.

\subsection{Reconstruction Task}

We train the meta-model $M_\theta$, reconstruct all videos in the test set to obtain $\hat {\mathcal{V}}$ as described in Figure~\ref{fig:inference}, and compute the Peak Signal-to-Noise Ratio (PSNR) and 3D Structural Similarity Index (SSIM3D) between the original video $ \mathcal{V}$ and the reconstruction $\hat{\mathcal{V}}$. We compare our time-resolved \textit{VidFuncta} against \textit{MedFuncta 2D} \cite{friedrich2025medfuncta}, which processes each frame individually, as well as its \textit{3D} variant trained on spatiotemporal chunks of size  $112\times112\times10$.
In addition to 1D latent modulations, we also implement \textit{Spatial Functa} \cite{bauer2023spatial}, which structures the latent modulations into a $4 \times 4\times 64$ grid, maintaining a comparable compression rate.
 We evaluate reconstruction quality when training our method on the \textit{mixed dataset}. For out-of-distribution (\textit{OOD}) experiments, we train on two datasets and run inference on the third. 

\begin{table}[t]
\caption{Reconstruction results of all comparing methods across the test set.}\label{tab1}
\centering
\begin{tabular}{l|cc|cc|cc}
\hline
 & \multicolumn{2}{c|}{\textbf{Cardiac}} & \multicolumn{2}{c|}{\textbf{Lung}} & \multicolumn{2}{c}{\textbf{Breast}} \\
 & SSIM3D &  PSNR  & SSIM3D &   PSNR & SSIM3D &  PSNR \\

\hline
\textit{MedFuncta 2D} & 90.8 $\pm$ 1 & 32.2 $\pm$ 1 & 82.4 $\pm$ 6   &29.5 $\pm$ 2 &   66.4 $\pm$ 5  &23.7 $\pm$ 1  \\
\textit{MedFuncta 3D} & 77.2 $\pm$  3& 27.6 $\pm$ 1 &  75.5 $\pm$ 8&  25.5 $\pm$ 3   & 48.8  $\pm$ 5 &18.1 $\pm$  3 \\
\textit{Spatial Functa} & 79.1 $ \pm$ 3 & 28.3 $\pm$ 2&  79.9 $\pm$ 7  &   28.5  $\pm$ 2  & 57.2  $\pm$ 4 & 22.2 $\pm$ 1   \\
\textit{VidFuncta} (Ours) &  \textbf{92.8} $\pm$ 2 & \textbf{34.2} $\pm$ 1&  \textbf{84.5} $\pm$ 6 & 30.0 $ \pm$ 3 & \textbf{71.1} $\pm$ 7 & \textbf{24.8} $\pm$ 1\\

\hline
Ours \textit{mixed dataset}& 84.3 $\pm$ 2 & 32.2 $\pm$ 1 & 82.7 $\pm$ 5 &\textbf{ 30.2} $\pm$ 2 &  68.9 $\pm$ 7& 24.3 $\pm$  2  \\
Ours \textit{OOD}  &68.0 $\pm$ 4 & 27.6 $\pm$ 2&  72.3 $\pm$ 5 & 27.6 $\pm$ 2 &51.5  $\pm$  5 & 21.2  $\pm$ 1 \\

\hline
\end{tabular}
\end{table}

\subsection{Downstream Tasks}

We test our model on three downstream tasks: ejection fraction prediction on cardiac ultrasound, B-line classification on lung ultrasound, and lesion classification on breast ultrasound. We evaluate performance across three input settings: the time-resolved representations $\phi=\{\phi_t\}_{t=1}^T$ alone, the video-specific vector  $v$  alone, and their combination. For the time-resolved inputs, we use a transformer encoder \cite{dosovitskiy2020image} with 2 heads and 4 layers. When combining with $v$, we append a linear embedding of $v$ to the sequence $\phi$. When using $v$  alone, we apply a 3-layer MLP with ReLU activations and dropout. We compare the performance on \textit{VidFuncta} modulation vectors with those from \textit{MedFuncta 2D} and \textit{Spatial Functa}. 
We additionally compare to convolutional video models, namely the  \textit{R(2+1)D} \cite{tran2018closer} architecture, and the 3D version of \textit{PocovidNet} \cite{born2020pocovid}.
For the regression task, we report the mean absolute error (MAE), root mean squared error (RMSE), and R$^2$ score. For binary classification tasks, we report the area under the receiver operating characteristic curve (AUROC), accuracy (ACC), and F1-score.

\section{Results and Discussion}

\subsection{Reconstruction Results}

\begin{figure}[t]
    \centering
    \includegraphics[width=0.92\linewidth]{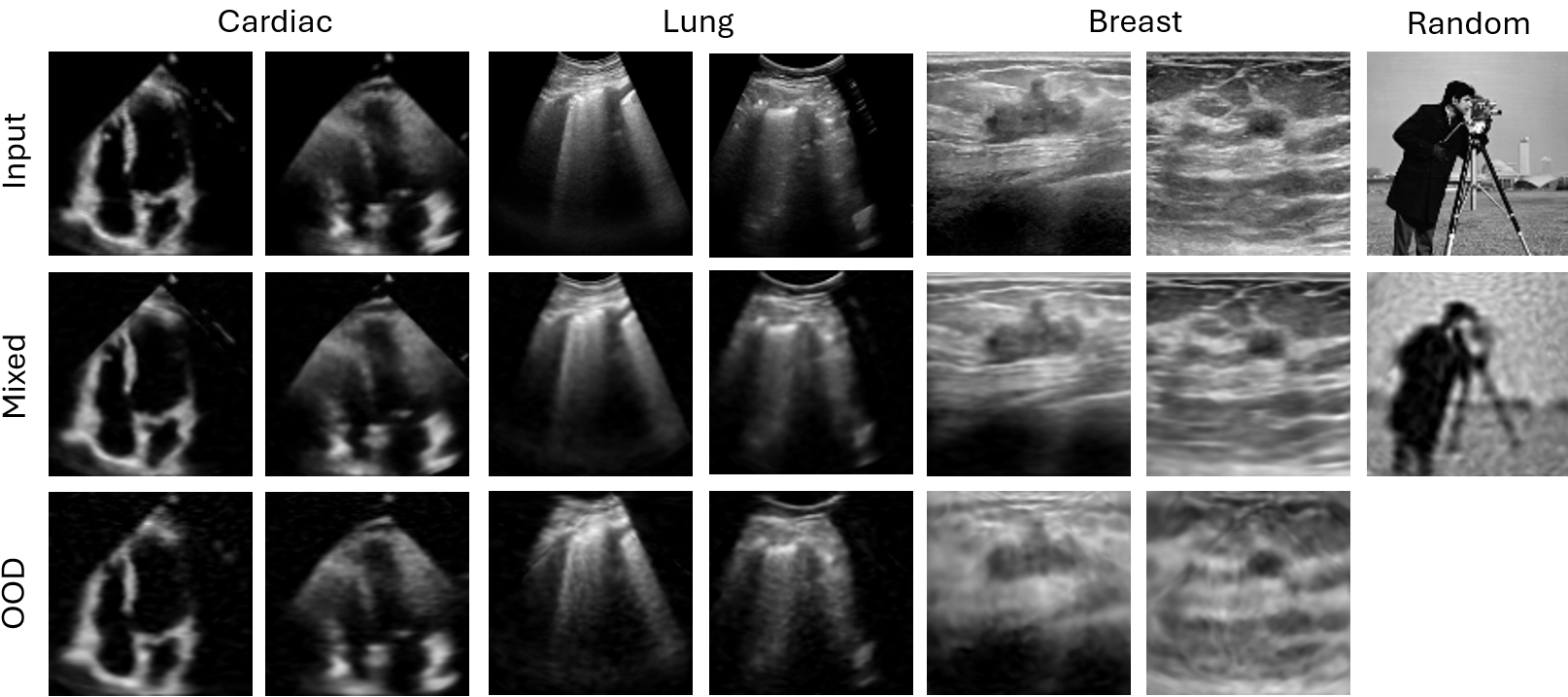}
    \caption{
    \textit{VidFuncta}  reconstructions on the \textit{mixed dataset}, as well as the \textit{OOD} results. The column "Random" shows the reconstruction of a natural image.}
    \label{fig:recon}
\end{figure}

\begin{figure}[t]
    \centering
    \includegraphics[width=\linewidth]{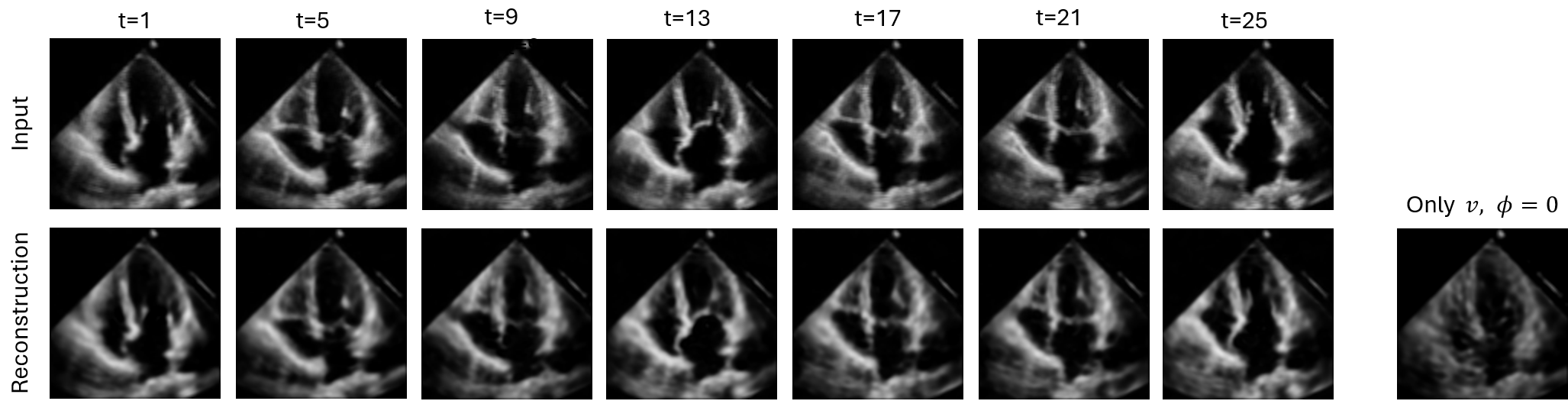}
    \caption{
    Series of reconstructed frames from a cardiac video using the dataset-specific \textit{VidFuncta} model, alongside the reconstruction using only $v$ while setting $\phi=0$.}
    \label{fig:series}
\end{figure}

Table~\ref{tab1} reports the mean $\pm$ standard deviation across the test set. Our autoregressive approach \textit{VidFuncta} achieves the best performance, outperforming both the frame-wise \textit{MedFuncta 2D} baseline and its \textit{3D} variant.  Using \textit{Spatial Functa} to structure modulation vectors reduces reconstruction quality, likely due to its shorter vector length $r$. Training on the \textit{mixed dataset} does not significantly degrade performance compared to training separate models per dataset, supporting the feasibility of a unified model across ultrasound modalities.
Figure~\ref{fig:recon} shows example reconstruction results from \textit{VidFuncta} for the mixed dataset, as well as the \textit{OOD} results. Reconstructions remain visually plausible in the \textit{OOD} setting, although the scores drop. Overall, reconstructing high-frequency details remains difficult. When tested on a natural image, as shown in Figure~\ref{fig:recon} on the right, the model produces ultrasound-like patterns while preserving key visual features, highlighting the potential for style transfer to unseen domains.
Figure~\ref{fig:series} visualizes a cardiac video sequence and its accurate reconstruction using \textit{VidFuncta}. On the right, we plot the reconstructed image using only the video-level modulations $v$ and setting $\phi_t=0$, which captures a summary of the entire sequence. Reconstructed videos and visualizations of $\phi$ are available in the project’s code repository. In Figure~\ref{fig:scatter} on the right, we show the t-SNE plot \cite{van2008visualizing} of the video-specific modulations $v$ from the \textit{mixed dataset}. The embeddings cluster clearly by modality, indicating that $v$ captures dataset-specific information.

\subsection{Results on the Downstream Tasks}

\begin{figure}[t]
    \centering
    \includegraphics[width=0.9\linewidth]{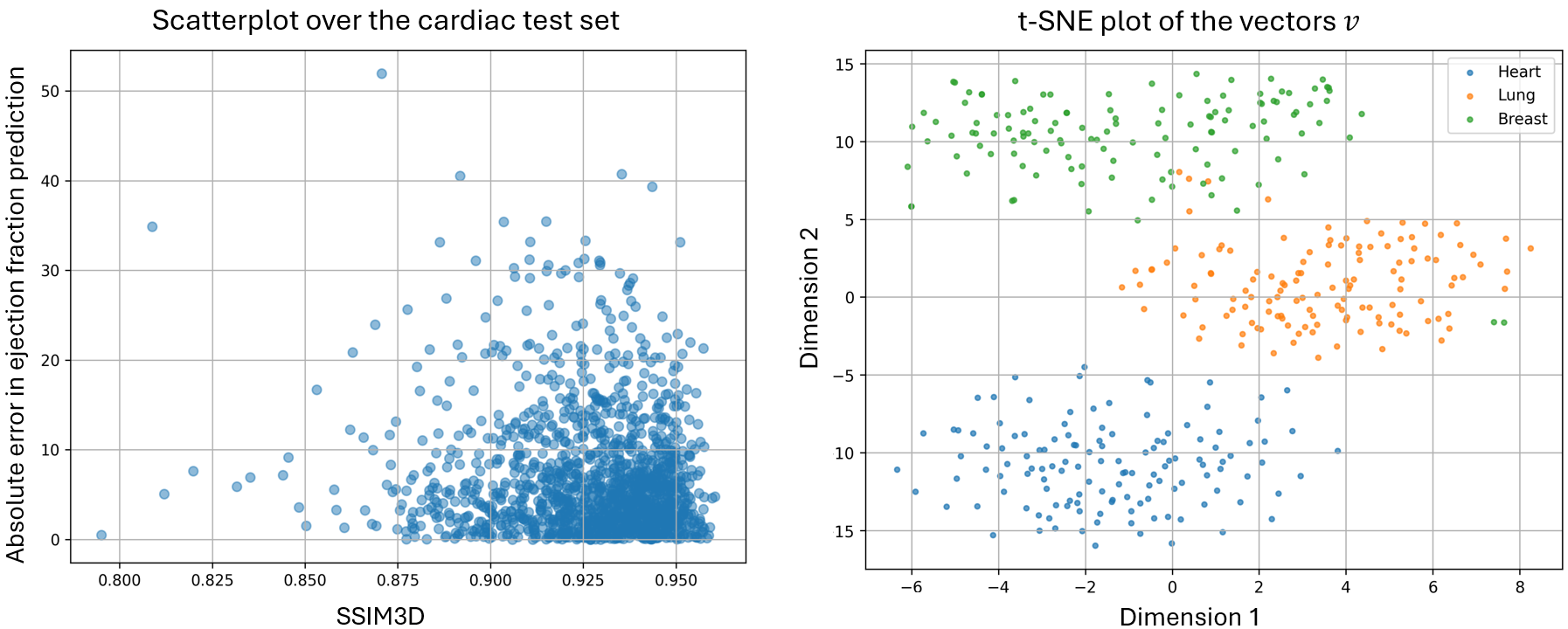}
    \caption{
    On the left, we plot the SSIM3D vs. absolute error in ejection fraction for \textit{VidFuncta} over the cardiac test set. On the right is the t-SNE plot of the vectors $v$ of the \textit{mixed dataset}, colored by modality. }
    \label{fig:scatter}
\end{figure}

Table~\ref{tab2} shows initial downstream regression and classification results. On the cardiac dataset, using only  $\phi=\{\phi_t\}_{t=1}^T$  yields the best performance, suggesting that temporal information is effectively captured in the sequence. 
While our setup has lower performance compared to convolutional baselines \textit{R(2+1)D} and \textit{PocovidNet}, evaluating \textit{PocovidNet} on reconstructed videos $\hat{\mathcal{V}}$ performs similarly as on $\mathcal{V}$, suggesting that task-relevant information is preserved during compression. 
We assume that latent modulations encode key features, but current downstream models cannot effectively extract them, as discussed in prior work \cite{bauer2023spatial,papa2024train}. Figure~\ref{fig:scatter} on the left shows that reconstruction quality of \textit{VidFuncta} on the cardiac test set does not correlate with the downstream performance. We experimented with \textit{Spatial Functa} to impose more structure on $\phi$, but found no performance gain. 
These results highlight the need for more structured and task-aligned approaches to extract $v$ and $\phi$.
For training 30 epochs with batch size 10, \textit{PocovidNet} requires 8.0 GB of memory and 4.5 hours, while \textit{VidFuncta} reduces this to 11 minutes and 0.35 GB.
For breast lesion classification, the model using both $v$ and $\phi$ performs best, comparable to \textit{PocovidNet} on both  $\mathcal{V}$ and $\hat{\mathcal{V}}$.
On the lung dataset,  while the convolutional models reach a high performance on both $\mathcal{V}$ and $\hat{\mathcal{V}}$,
the performance of all \textit{Functa} approaches remains limited.
We observe overfitting on the training set, highlighting the need to improve downstream architectures and generalization techniques on the modulations.\\

\begin{table}[t]
\caption{Mean of the downstream test performance across 5 folds. A more detailed table including the standard deviation is provided in the code repository.}\label{tab2}
\centering
\begin{tabular}{l|p{0.07\textwidth} p{0.08\textwidth}l|p{0.065\textwidth}p{0.065\textwidth}l|p{0.065\textwidth}p{0.065\textwidth}l}
\hline

& \multicolumn{3}{c|}{\textbf{Cardiac}} & \multicolumn{3}{c|}{\textbf{Lung }}  & \multicolumn{3}{c}{\textbf{Breast}} \\
& MAE & RMSE & R2-Score & Acc & F1 & AUROC &  Acc & F1 & AUROC \\
\hline
\textit{MedFuncta 2D}& 6.82  & 9.34 & 0.39   &  64.6 &72.7 & 65.1  & 71.0 &77.4 & 69.9 \\

\textit{Spatial Functa }& 7.72
&10.86  & 0.41
& 63.4 
& 71.2  
& 65.8

& 70.1 & 77.3 &  81.7 
\\

\textit{VidFuncta} only $\phi $ &  6.11 &8.35 & 0.56  & 62.5  & 70.3  & 65.6   & 64.6  & 68.1 & 63.0 \\

\textit{VidFuncta} only $v$  &9.81 &12.10& 0.15  &63.7 &72.8 &64.2 &72.6 &77.9 &74.3\\

\textit{VidFuncta} $v \times  \phi $ & 6.28  &8.72   & 0.54 
& 62.6 &70.3 & 65.1& 76.4  & 82.9 & 77.4 \\
\hline
\textit{R(2+1)D} on $\mathcal{V}$ & 4.87  & 6.52 & 0.72   & 77.9  & 82.0  & 83.9 & 60.0 & 67.0   & 64.8   \\

\textit{PocovidNet} on $\mathcal{V}$  & 4.34   &5.84   &   0.77&   86.7 & 88.2 & 93.1 & 76.1 & 80.1   & 82.3 \\

\textit{PocovidNet}  on $\hat{\mathcal{V}}$ &4.60   & 6.24 & 0.75  & 83.0 &  85.1  & 90.2   & 73.4 & 80.1 & 72.6 \\
\hline
\end{tabular}
\end{table}

\normalsize

\section{Conclusion}

We present \textit{VidFuncta}, a novel framework for time-resolved compressed neural representations of ultrasound videos, enabling high-quality reconstructions and downstream tasks on sequences of 1D modulation vectors. Our method outperforms 2D and 3D baselines, supports multiple ultrasound datasets within a single unified model, and generalizes well to out-of-distribution data. Downstream training time and memory use is reduced by roughly 25× compared to convolution-based approaches.
Some limitations remain: High-frequency details are poorly preserved, lowering reconstruction scores.
Future work will explore alternative architectures such as \textit{WIRE} and \textit{FINER} activations \cite{liu2024finer,saragadam2023wire} to address this issue.
 We will further explore the relationship between compression rate and reconstruction quality. 
Current downstream models struggle to fully leverage the compressed modulation vectors; improved structuring of the modulations may enhance task-specific performance. 
Overall, this work introduces a new direction for ultrasound video compression and analysis, and opens the door to a wide range of applications such as domain generalization and style transfer.

\begin{credits}
\subsubsection{\ackname} This work was supported by the Swiss National Science Foundation (Grant No. P500PT$\_$222349). We thank Prof. Tina Kapur for the valuable discussions and for providing access to the lung ultrasound dataset.

\subsubsection{\discintname}
The authors have no competing interests to declare that are
relevant to the content of this article.
\end{credits}

%
%
\bibliographystyle{splncs04}
\bibliography{bibilography.bib}

\begin{thebibliography}{10}
\providecommand{\url}[1]{\texttt{#1}}
\providecommand{\urlprefix}{URL }
\providecommand{\doi}[1]{https://doi.org/#1}

\bibitem{asgari2024can}
Asgari-Targhi, A., Ungi, T., Jin, M., Harrison, N., Duggan, N., Duhaime, E., Goldsmith, A., Kapur, T.: Can crowdsourced annotations improve ai-based congestion scoring for bedside lung ultrasound? In: International Conference on Medical Image Computing and Computer-Assisted Intervention. pp. 580--590. Springer (2024)

\bibitem{baloescu2020automated}
Baloescu, C., Toporek, G., Kim, S., McNamara, K., Liu, R., Shaw, M.M., McNamara, R.L., Raju, B.I., Moore, C.L.: Automated lung ultrasound b-line assessment using a deep learning algorithm. IEEE Transactions on Ultrasonics, Ferroelectrics, and Frequency Control  \textbf{67}(11),  2312--2320 (2020)

\bibitem{bauer2023spatial}
Bauer, M., Dupont, E., Brock, A., Rosenbaum, D., Schwarz, J.R., Kim, H.: Spatial functa: Scaling functa to imagenet classification and generation. arXiv preprint arXiv:2302.03130  (2023)

\bibitem{born2020pocovid}
Born, J., Br{\"a}ndle, G., Cossio, M., Disdier, M., Goulet, J., Roulin, J., Wiedemann, N.: Pocovid-net: automatic detection of covid-19 from a new lung ultrasound imaging dataset (pocus). arXiv preprint arXiv:2004.12084  (2020)

\bibitem{chen2022videoinr}
Chen, Z., Chen, Y., Liu, J., Xu, X., Goel, V., Wang, Z., Shi, H., Wang, X.: Videoinr: Learning video implicit neural representation for continuous space-time super-resolution. In: Proceedings of the IEEE/CVF Conference on Computer Vision and Pattern Recognition. pp. 2047--2057 (2022)

\bibitem{dosovitskiy2020image}
Dosovitskiy, A., Beyer, L., Kolesnikov, A., Weissenborn, D., Zhai, X., Unterthiner, T., Dehghani, M., Minderer, M., Heigold, G., Gelly, S., et~al.: An image is worth 16x16 words: Transformers for image recognition at scale. arXiv preprint arXiv:2010.11929  (2020)

\bibitem{dupont2022data}
Dupont, E., Kim, H., Eslami, S., Rezende, D., Rosenbaum, D.: From data to functa: Your data point is a function and you can treat it like one. arXiv preprint arXiv:2201.12204  (2022)

\bibitem{dupont2022coin++}
Dupont, E., Loya, H., Alizadeh, M., Goli{\'n}ski, A., Teh, Y.W., Doucet, A.: Coin++: Neural compression across modalities. arXiv preprint arXiv:2201.12904  (2022)

\bibitem{eid2025rapidvol}
Eid, M.C., Yeung, P.H., Wyburd, M.K., Henriques, J.F., Namburete, A.I.: Rapidvol: Rapid reconstruction of 3d ultrasound volumes from sensorless 2d scans. In: 2025 IEEE 22nd International Symposium on Biomedical Imaging (ISBI). pp.~1--5. IEEE (2025)

\bibitem{friedrich2025medfuncta}
Friedrich, P., Bieder, F., Cattin, P.C.: Medfuncta: Modality-agnostic representations based on efficient neural fields. arXiv preprint arXiv:2502.14401  (2025)

\bibitem{gu2022representing}
Gu, A.N., Abolmaesumi, P., Luong, C., Yi, K.M.: Representing 3d ultrasound with neural fields. In: Medical Imaging with Deep Learning (2022)

\bibitem{guo2025efficient}
Guo, J., Wu, Y., Kaimakamis, E., Petmezas, G., Papageorgiou, V.E., Maglaveras, N., Katsaggelos, A.K.: Efficient lung ultrasound severity scoring using dedicated feature extractor. arXiv preprint arXiv:2501.12524  (2025)

\bibitem{howard2020improving}
Howard, J.P., Tan, J., Shun-Shin, M.J., Mahdi, D., Nowbar, A.N., Arnold, A.D., Ahmad, Y., McCartney, P., Zolgharni, M., Linton, N.W., et~al.: Improving ultrasound video classification: an evaluation of novel deep learning methods in echocardiography. Journal of medical artificial intelligence  \textbf{3}, ~4 (2020)

\bibitem{Yurong2023Self-supervised}
Hu, Yurong, e.a.: Self-supervised learning to predict ejection fraction using motion-mode images. 1st Workshop on Machine Learning and Global Health (ICLR 2023)  (2023)

\bibitem{huang2023breast}
Huang, Y., Hu, H., Zhu, Y., Xu, Y.: Breast lesion diagnosis using static images and dynamic video. In: 2023 IEEE 20th International Symposium on Biomedical Imaging (ISBI). pp.~1--5. IEEE (2023)

\bibitem{kim2021artificial}
Kim, Y.H.: Artificial intelligence in medical ultrasonography: driving on an unpaved road. Ultrasonography  \textbf{40}(3), ~313 (2021)

\bibitem{lin2022new}
Lin, Z., Lin, J., Zhu, L., Fu, H., Qin, J., Wang, L.: A new dataset and a baseline model for breast lesion detection in ultrasound videos. In: International Conference on Medical Image Computing and Computer-Assisted Intervention. pp. 614--623. Springer (2022)

\bibitem{liu2024finer}
Liu, Z., Zhu, H., Zhang, Q., Fu, J., Deng, W., Ma, Z., Guo, Y., Cao, X.: Finer: Flexible spectral-bias tuning in implicit neural representation by variable-periodic activation functions. In: Proceedings of the IEEE/CVF Conference on Computer Vision and Pattern Recognition. pp. 2713--2722 (2024)

\bibitem{lucassen2023deep}
Lucassen, R.T., Jafari, M.H., Duggan, N.M., Jowkar, N., Mehrtash, A., Fischetti, C., Bernier, D., Prentice, K., Duhaime, E.P., Jin, M., et~al.: Deep learning for detection and localization of b-lines in lung ultrasound. IEEE journal of biomedical and health informatics  \textbf{27}(9),  4352--4361 (2023)

\bibitem{van2008visualizing}
Van~der Maaten, L., Hinton, G.: Visualizing data using t-sne. Journal of machine learning research  \textbf{9}(11) (2008)

\bibitem{ouyang2019echonet}
Ouyang, D., He, B., Ghorbani, A., Lungren, M.P., Ashley, E.A., Liang, D.H., Zou, J.Y.: Echonet-dynamic: a large new cardiac motion video data resource for medical machine learning. In: NeurIPS ML4H Workshop: Vancouver, BC, Canada. vol.~5 (2019)

\bibitem{papa2024train}
Papa, S., Valperga, R., Knigge, D., Kofinas, M., Lippe, P., Sonke, J.J., Gavves, E.: How to train neural field representations: A comprehensive study and benchmark. In: Proceedings of the IEEE/CVF Conference on Computer Vision and Pattern Recognition. pp. 22616--22625 (2024)

\bibitem{perez2018film}
Perez, E., Strub, F., De~Vries, H., Dumoulin, V., Courville, A.: Film: Visual reasoning with a general conditioning layer. In: Proceedings of the AAAI conference on artificial intelligence. vol.~32 (2018)

\bibitem{rumack2023diagnostic}
Rumack, C.M., Levine, D.: Diagnostic ultrasound E-book. Elsevier Health Sciences (2023)

\bibitem{saragadam2023wire}
Saragadam, V., LeJeune, D., Tan, J., Balakrishnan, G., Veeraraghavan, A., Baraniuk, R.G.: Wire: Wavelet implicit neural representations. In: Proceedings of the IEEE/CVF Conference on Computer Vision and Pattern Recognition. pp. 18507--18516 (2023)

\bibitem{sitzmann2020implicit}
Sitzmann, V., Martel, J., Bergman, A., Lindell, D., Wetzstein, G.: Implicit neural representations with periodic activation functions. Advances in neural information processing systems  \textbf{33},  7462--7473 (2020)

\bibitem{stewart2020trends}
Stewart, K.A., Navarro, S.M., Kambala, S., Tan, G., Poondla, R., Lederman, S., Barbour, K., Lavy, C.: Trends in ultrasound use in low and middle income countries: a systematic review. International Journal of Maternal and Child Health and AIDS  \textbf{9}(1), ~103 (2020)

\bibitem{tran2018closer}
Tran, D., Wang, H., Torresani, L., Ray, J., LeCun, Y., Paluri, M.: A closer look at spatiotemporal convolutions for action recognition. In: Proceedings of the IEEE conference on Computer Vision and Pattern Recognition. pp. 6450--6459 (2018)

\bibitem{wang2021deep}
Wang, Y., Ge, X., Ma, H., Qi, S., Zhang, G., Yao, Y.: Deep learning in medical ultrasound image analysis: a review. Ieee Access  \textbf{9},  54310--54324 (2021)

\bibitem{wiedemann2025covid}
Wiedemann, N., de~Korte-De~Boer, D., Richter, M., van~de Weijer, S., Buhre, C., Eggert, F.A., Aarnoudse, S., Grevendonk, L., R{\"o}ber, S., Remie, C.M., et~al.: Covid-blues-a prospective study on the value of ai in lung ultrasound analysis. IEEE Journal of Biomedical and Health Informatics  (2025)

\end{thebibliography}
%

\end{document}